\documentclass{elsart}

\usepackage{amssymb}
\usepackage{graphicx}
\newcommand{\be}{\begin{eqnarray}}
\newcommand{\ee}{\end{eqnarray}}

\begin{document}

\begin{frontmatter}


 \title{Intermediate mass dileptons in relativistic nuclear collisions}
 \author{Charles Gale}, 
 \ead{gale@physics.mcgill.ca}
 \ead[url]{www.physics.mcgill.ca/$\sim$gale}
 \author{Simon Turbide}
 \ead{simon.turbide@mail.mcgill.ca}

 \address{Department of Physics, McGill University\\ 3600 University Street, Montreal, QC, Canada H3A 2T8}





\begin{abstract}

We consider the production of lepton pairs with intermediate masses, in relativistic nuclear collisions. Thermal sources are briefly discussed, as well as a newer production mechanism which involves jet-plasma interactions. Estimates for high $p_T$ lepton pairs are provided.
\end{abstract}

\end{frontmatter}

\vspace*{-5mm}
\section{Introduction}
\vspace*{-5mm}

A variety of observables characterizes the final state of relativistic nuclear collisions. A plethora of hadronic particle species can indeed be measured, and precious information about the breakup conditions, flow profile, and relative populations can be acquired. Penetrating probes, however, carry complementary information: they offer the possibility of inferring what the local conditions were at early times in the reaction history. Electromagnetic observables, real photons and lepton pairs, belong to this category: they will be emitted from the very first instant of the nuclear collision - when the first nucleons feel each other -  to the very last moment when the strongly interacting matter decouples and flies toward the detectors. Therefore, any quantitative assessment of the electromagnetic signal needs to be based on a reliable estimate of the emissivity in each phase of the evolving system. Only then will the signal to background ratio be convincingly determined. In the context of the search for the  quark-gluon plasma,  the measurement of dileptons at intermediate invariant mass ($m_\phi < M < m_{J/\psi}$) was among one of the early suggestions \cite{shu}. A combination of kinematical factors makes this region especially interesting: it was thought to be high enough in invariant mass for the usual thermal hadronic contributions to be sub-dominant, and low enough for hard QCD processes not to be overwelming. A confirmation of this elegant and simple idea   however requires a careful consideration of the details of the different emitting processes.

\vspace*{-5mm}
\section{Thermal dileptons}
\vspace*{-5mm}

\begin{figure}[h!]
\begin{center}
\includegraphics*[width=6cm]{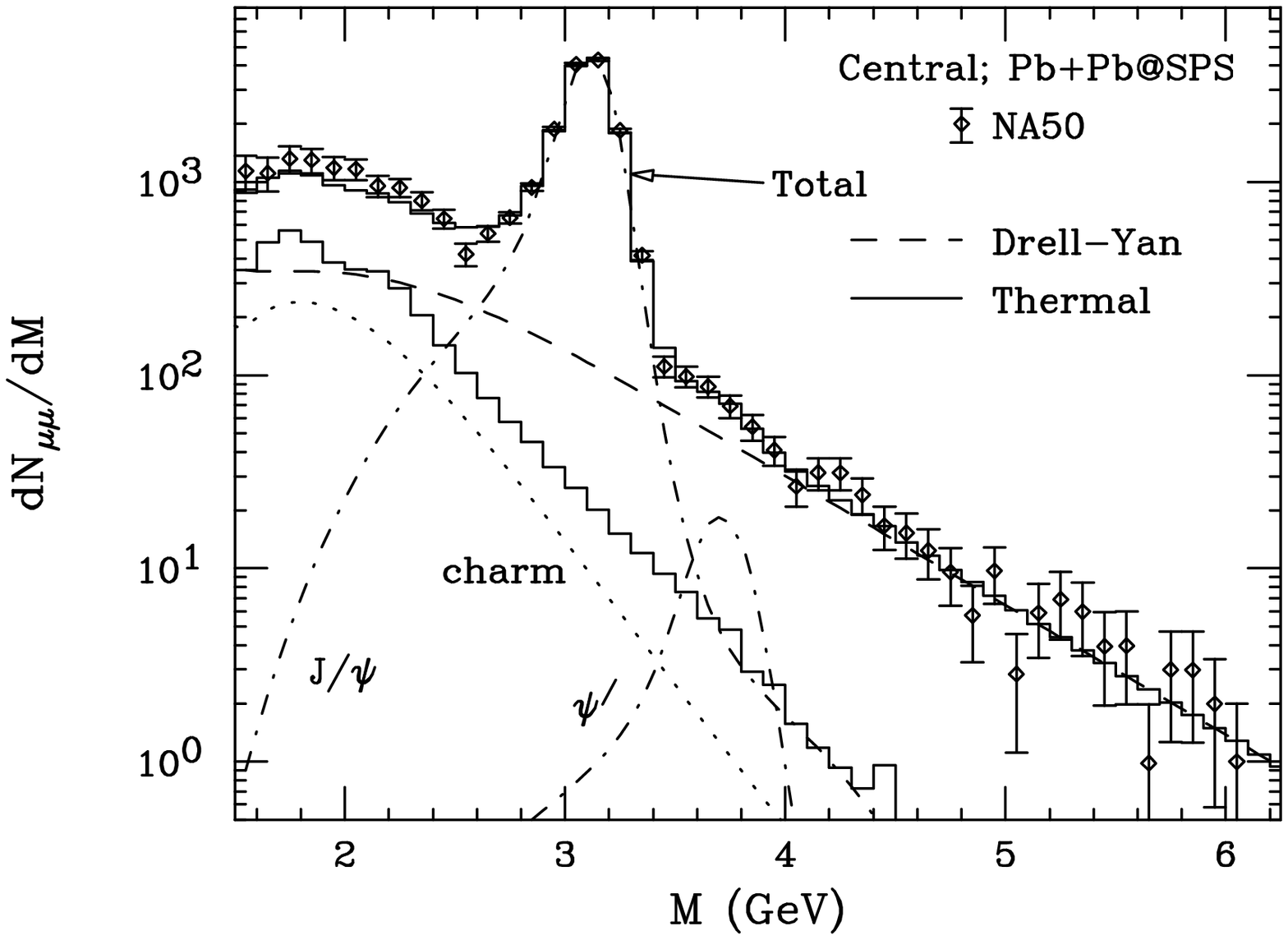}
\includegraphics*[width=6cm]{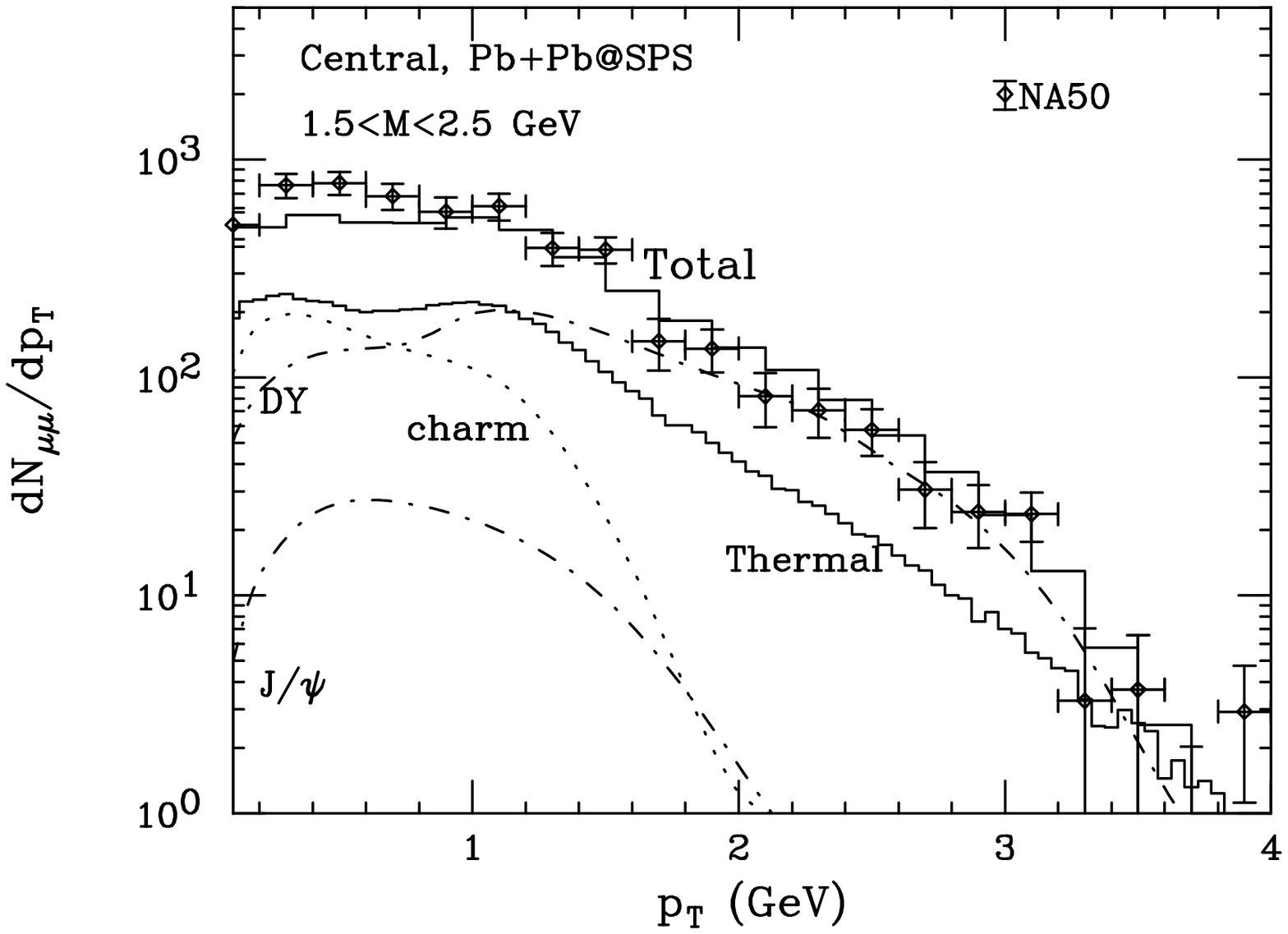}
\caption{The dimuon invariant mass and momentum spectra. The represented sources are Drell-Yan (DY), correlated charm decay, and thermal (quark-gluon plasma and hadron gas) \cite{Kvasnikova:2001zm}. The full curve histogram is the net signal, after correcting for detector acceptance, resolution and efficiency. The data is from the NA50 collaboration \cite{NA50}.}
\label{fig1}
\end{center}
\end{figure}

Intermediate mass measurements have been realized within a vibrant experimental program, going on for roughly two decades. Indeed, considerable interest was generated by the fact that an excess over sources expected from $pA$ measurements has been confirmed early in the relevant mass region by the Helios/3 \cite{helios} and NA50 \cite{NA50} collaborations. This excess could perhaps be interpreted in terms of an anomalously high $c \bar{c}$ abundance, which would then in turn manifest itself through the correlated measurement of the semileptonic decay of open charm mesons \cite{shor}. Alternatively, thermal lepton pairs need to be ruled out as a viable scenario before other sources can be invoked. Typical calculations of lepton pair emission rates involve effective Lagrangians, whose parameters are fitted to hadronic and electromagnetic decay widths \cite{Rapp:1999qu}. However, when extrapolating to harder energy scales, caution should be exerted as off-shell effects might become large \cite{Gao:1997vm}. Fortunately, a way out of this theoretical ambiguity is provided by the abundance of $e^+ e^- \to {\rm hadrons}$ data which exists, and covers precisely the required range in invariant mass \cite{dol}. These can then be used to construct rates, which can be integrated assuming a space-time evolution scenario for the colliding system. The result of one such exercise is shown in Figure \ref{fig1}.
It appears, as it did in an earlier analysis of the Helios/3 data \cite{Li:1998xn}, that the thermal sources are large enough for the data not to require a large quark-gluon plasma contribution, nor a large enhancement of $c \bar{c}$ pairs. The detailed analysis does depend on the details of the space-time modeling, but this conclusion is fairly robust and was also reached by other theoretical studies of a similar nature \cite{na50:analyses}. A direct measurement of the strangeness content could contribute to settle this issue. In this respect, a breakthrough was recently realized by the NA60 collaboration. The enhancement in the intermediate mass sector was indeed confirmed by their measurement, and shown to be consistent with that reported by NA50. Moreover, NA60 can now assert that this excess is not due to open charm enhancement \cite{ad}. The precise nature of this data will undoubtedly constitute fertile grounds for theoretical analyses, some of which have begun \cite{VHR}. We now turn to RHIC energies, where dilepton measurements are eagerly anticipated and where the higher colliding energies might reveal additional sources.

\vspace*{-5mm}
\section{Dileptons from jet-plasma interactions}
\vspace*{-5mm}

One of the main experimental results from the RHIC program has been the observation of spectacular jet quenching in central collisions of relativistic nuclei \cite{gyulassy}. This signature is taken to reflect a strong interaction of the leading parton with a colored medium. If these jet-plasma interactions are so efficient in altering the jet properties, they should also be able to produce electromagnetic radiation. It has been shown that the intermediate to high $p_T$ photon spectrum can indeed contain a sizeable contribution of this nature \cite{fms,tgjm}. In principle these new channels should also be effective in generating lepton pairs; the question is whether they can shine through the background which is not the same as in the case of real photons. 

\begin{figure}[h!]
\begin{center}
\includegraphics*[width=10cm]{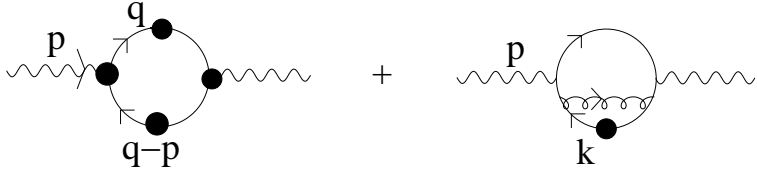}
\caption{The resummed photon self-energy diagram. Full lines are quark fields, and curly lines are gluons. The heavy dots signify effective vertices and propagators. }
\label{pi_phot}
\end{center}
\end{figure}

A large background for the production of intermediate and high mass lepton pairs at RHIC energies has been, like at the SPS, that associated with the correlated measurement of semileptonic charm decays: $D \bar{D} \to \ell^+ \ell^- X$. Only here, this irreducible contribution is indeed expected to dominate over the thermal yields \cite{Gavin:1996bx}.  Similarly, dileptons from Drell-Yan \cite{DY}, thermal dileptons, and dileptons from the fragmentation of jets have to be considered, as these will compete with the lepton pairs coming from the jet-plasma interactions. Here, we therefore examine the dilepton production from jets and we include the effect of energy-loss. We assume here that jets loose their energy via induced gluon bremsstrahlung, and we use the approach of AMY \cite{amy}. This discussion follows that in Ref. \cite{tgsf}. Jets will be defined here as partons produced initially with $p_T \gg$ 1 GeV. The total dilepton production could be influenced by the choice of this cutoff: in order to avoid such sensitivity we limit this study to high momentum dileptons. In high-temperature field theory, the Hard Thermal Loops (HTL) formalism will take care of the divergences that will appear if only diagrams of low-order topologies are used \cite{KG}, and the production rate of lepton pairs is given in terms of the imaginary part of the finite-temperature, retarded, in-medium photon self-energy \cite{rate}
\be
E_+ E_- \frac{d^6 R}{d^3 p_+ d^3 p_-} = \frac{2 e^2}{(2 \pi)^6} \frac{1}{p^4} L^{\mu \nu} {\rm Im} \Pi^{\rm R}_{\mu \nu} (k) \frac{1}{{\rm e}^{\beta p_0} - 1}\parbox{16ex}{\ } 
\ee
In the above equation the lepton tensor is simply defined in terms of the lepton four-momenta and mass: $L^{\mu \nu} = p^\mu_+ p^\nu_- + p^\mu_- p^\nu_+ - g^{\mu \nu} \left( p_+ \cdot p_- + m^2 \right)$, and $p^\mu$ is the photon four-momentum (real or virtual).
 The leading order HTL-resummed photon self-energy is shown in the left hand side of Figure \ref{pi_phot}. The heavy dots indicate a resummed propagator or effective vertex. The second diagram is needed in order to fulfill the Ward identity.
Since we are interested only in high-momentum dileptons, vertex corrections will play no role as at least one propagator is guaranteed to be hard. It is limit, the self-energy becomes
\be
\label{pi_1}
  \Pi^{\mu}_{\mu}(p) = 3e^2\sum_f\left(\frac{e_f}{e}\right)^2T\sum_n\int 
  \frac{d^3q}{(2\pi)^3}\mbox{Tr}\left[\gamma^{\mu}S_D(q)\gamma_{\mu}S(q-p) 
  \right]
\ee
with the sum running over the usual Matsubara frequencies. This treatment includes 
the leading order effect, and some next-to leading order corrections in $g_s$.
In order to have a complete next-to leading order production rate in the region $|\overrightarrow{p}|\gg T$, contributions like bremsstrahlung need also to be included. We will not discuss these here, but their importance can be estimated from those in the case of real photons \cite{tgjm} \footnote{This explicit calculation has since been completed. Results that also include a realistic hydrodynamic modeling are forthcoming.}. 
The dressed fermion propagator, in Euclidean space 
($\not\!q=-iq_0\gamma^0+\vec{q}\cdot\hat{\gamma}$), is given
  by~\cite{KG} 
\be
S_D(q)=\frac{1}{\not\!q+\Sigma}=\frac{\gamma^0-\hat{\gamma}\cdot \hat{\bf q}}{2D_+(q)}+\frac{\gamma^0+\hat{\gamma}\cdot \hat{\bf q}}{2D_-(q)}
\ee
where
\be
\label{D_def}
D_{\pm}(q)=-iq_0\pm|\overrightarrow{q}|+A\pm B
\ee
The terms $A$ and $B$ describe the quark self-energy  
\be
\Sigma=A\gamma^0+B\hat{\gamma}\cdot \hat{\bf q}.
\ee

\begin{figure}[h!]
\begin{center}
\includegraphics*[angle=-90,width=8cm]{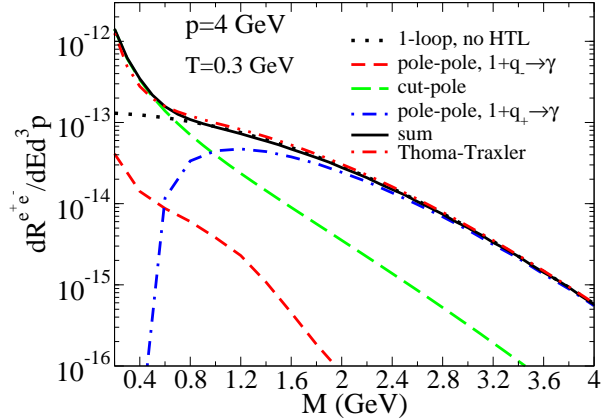}
\caption{The production rate of dileptons with momentum $p = 4$ GeV, with $T =$ 0.3 GeV, and using $\alpha_s = 0.3$. The dotted line represents the Born term, and the other different contribution are explained in \cite{tgsf}. The double-dot dashed line is the result of Ref. \cite{TT}.}  
\label{tt}
\end{center}
\end{figure}

\begin{figure}[h!]
\begin{center}
\includegraphics*[angle=0,width=10cm]{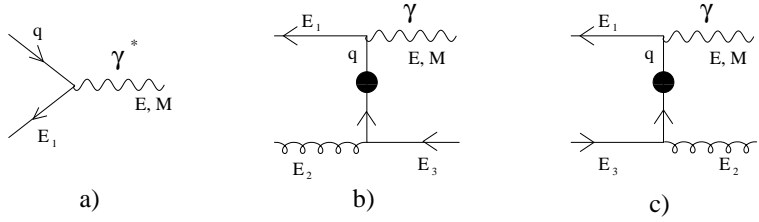}
\caption{The physical processes that correspond to taking the high-momentum limit of the imaginary part of Figure \ref{pi_phot}. The labels $E, M$ refer to the energy and the invariant mass, respectively.}  
\label{feyn2}
\end{center}
\end{figure}

\begin{figure}[h!]
\begin{center}
\includegraphics*[angle=-90,width=6.5cm]{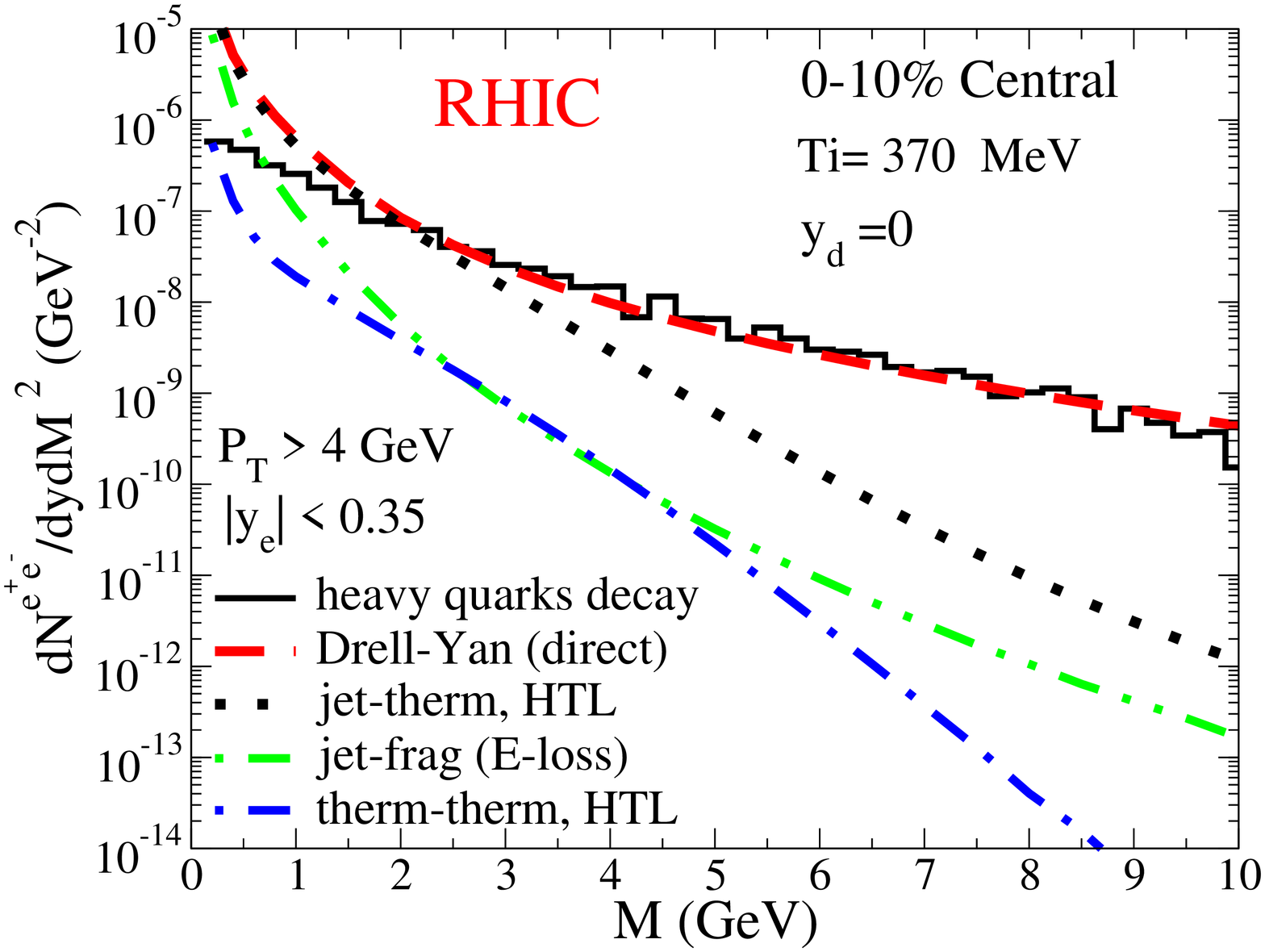}
\includegraphics*[angle=-90,width=6.5cm]{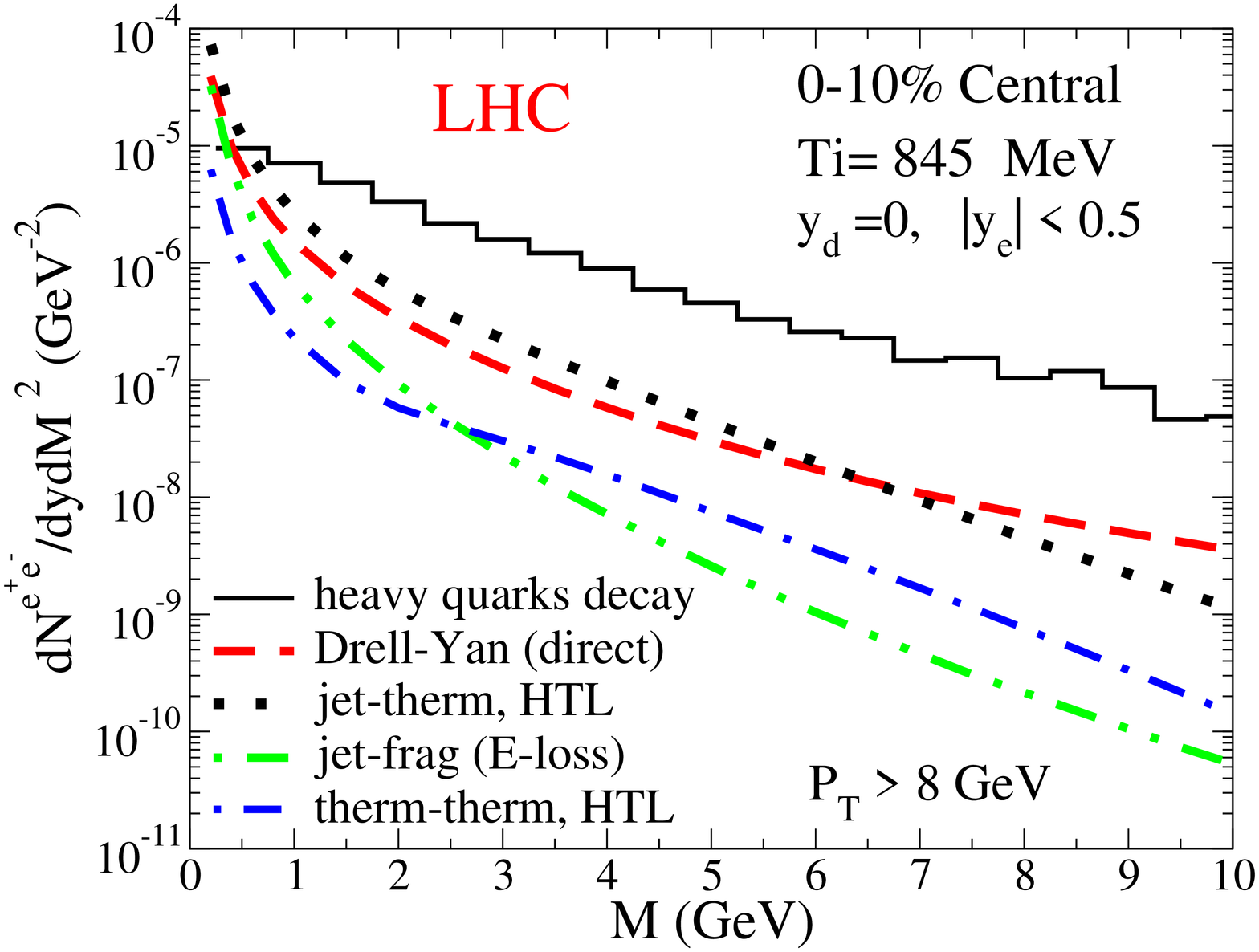}
\caption{The different sources of high $p_T$ dileptons in central Au + Au collisions at RHIC (left panel) and at the LHC (right panel). The solid curves represent the semileptonic decay of heavy quarks; the dashed lines: direct Drell-Yan contribution, dotted lines: jet-thermal interactions; double dotted-dashed lines: jet-fragmentation contribution; dotted-dashed lines: thermal dilepton production.}  
\label{result1}
\end{center}
\end{figure}

\begin{figure}[h!]
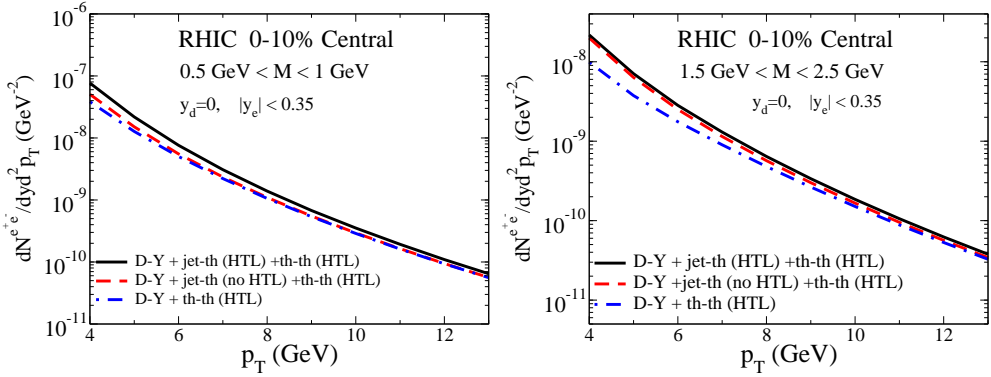

\begin{center}
\includegraphics*[angle=0,width=6.5cm]{dilep_rhic_sum_K.eps}
\includegraphics*[angle=0,width=6.5cm]{dilep_dptsum_rhic_M15_25_K.eps}
\caption{Transverse momentum distribution of dileptons, integrated in two different invariant mass ranges: 0.5 $< M <$ 1 GeV (left panel), and 1.5$ < M < $2.5 GeV (right panel), for central Au + Au collisions at RHIC. The solid lines: sum of DY, jet-thermal and purely thermal contributions; dashed lines: sum of DY, jet-thermal (without HTL) and purely thermal; dotted-dashed lines: sum of DY and thermal production.} 
\label{result2}
\end{center}
\end{figure}

In the HTL approximation, one obtains
\be
\label{A_B_def}
  A = \frac{m_F^2}{|\vec{q}|}Q_0\left(\frac{iq_0}{|\vec{q}|}\right),
  \quad B = -\frac{m_F^2}{|\vec{q}|}Q_1\left(\frac{iq_0}{|\vec{q}|}\right)
\ee
where $m_F=g_s T/\sqrt{6}$ is the effective quark mass induced by the thermal 
medium and the $Q_n$ are Legendre functions of the second kind. 
The bare propagator $S(q)$ is the same as $S_D(q)$, with $D_{\pm}$ 
replaced by
\be
  d_{\pm}(q)=-iq_0\pm|\vec{q}|
\ee
In order to make contact with the phase-space distribution associated with a hard jet, we have kept this function arbitrary, possibly non-thermal, and used the machinery of relativistic kinetic theory. This procedure however leaves us with a cutoff-dependence which we fortunately find to be weak \cite{tgsf}. Figure \ref{tt} contains a comparison of our calculation with that of Thoma and Traxler, who worked in the thermal limit \cite{TT}. 
The different processes correspond to the different channels in Figure \ref{feyn2} \cite{tgsf}. Our evaluation agrees well with that of \cite{TT}, and we now go on to the next step.
Turning to jet-thermal dilepton production, we use the same time-evolution scenario as that in our evaluation of the real photon spectrum  \cite{tgjm}. We calculate the Drell-Yan emission to  ${\mathcal O} (\alpha_s)$, which is the leading order non-vanishing result at finite $p_T$ of the lepton pair. We also need to take into account virtual photon bremsstrahlung from jets. The net DY yield is the sum of the direct and bremsstrahlung contributions \cite{dy2}. The decay of open charm and bottom mesons will play a large role, as already discussed. The correlated decays are calculated with the techniques of Ref. \cite{Mangano}. Further note that the collisional energy loss \cite{mt} of the heavy quarks has not been taken into account, and that the curves associated with them here will thus be upper limits. Adding all of this up, we get the results shown in Figure \ref{result1}. As expected, heavy quark decays dominate over most of the kinematical range, and significant progress can be made if a direct measurement (along the lines of what was done by NA60 at the SPS) is possible. However, some insight can be gained by looking at relative strengths. Indeed, the jet-thermal contribution can exceed - depending on the invariant mass - the purely thermal source by roughly an order of magnitude, as was the case for real photons \cite{tgjm}. At RHIC, for the intermediate masses relevant to this discussion, the jet-thermal source is as large as DY and heavy quark decay: clearly it needs to be part of the discussion of quark-gluon plasma signatures. 
It is also revealing to look at the dilepton spectrum in terms of $p_T$, for a given invariant mass window. The results for RHIC, in two regions of $M$, are shown in Figure \ref{result2}.
The left panel tells us about the important of HTL resummation in the low invariant mass window, and that the absence of any jet-thermal interactions would reduce the yield by a factor $\sim$2 at $p_T$ = 4 GeV. In the intermediate mass range (right panel), the interactions of jets with the plasma clearly play an important role.

\vspace*{-5mm}

\section{Conclusion}
\vspace*{-5mm}

The recent NA60 data at CERN SPS energies have shown that the excess observed at intermediate dilepton invariant masses could not be attributed to charm enhancement. This leaves the thermal sources very much in the running for the interpretation of data in this region, but complications that are germane to the nucleus-nucleus environment need to be quantitatively understood \cite{dy2}.  At RHIC energies, jet-thermal interactions will also contribution to the intermediate mass region. These sources need to be included in realistic estimates. 

\vspace*{-5mm}
\section*{Acknowledgments} 
\vspace*{-5mm}
We are grateful to all the collaborators that participated in obtaining the results shown here. 
This work was supported in part by the Natural Sciences and Engineering Research Council of Canada, and in part by the Fonds Nature et Technologies du Qu\'ebec.

\end{document}